\documentclass[twocolumn,prl,aps,citeautoscript,showpacs,footinbib,amsmath,amssymb,superscriptaddress,longbibliography]{revtex4-1}

\usepackage{graphicx}
\usepackage[utf8]{inputenc}
\usepackage{amsmath}
\usepackage{braket}
\usepackage{lmodern}
\usepackage[normalem]{ulem}
\usepackage{bbold}
\usepackage[T1]{fontenc}
\usepackage{xcolor}
\usepackage{siunitx}
\usepackage{comment}
\usepackage{hyperref}
\hypersetup{pdfnewwindow=true,allcolors=blue,colorlinks=true}
\usepackage{lineno}

\begin{document}
\title{Atomic-scale Dzyaloshinskii-Moriya-modified Yoshimori spirals\\
in Fe double layer on Ir(110)}

\author{Timo~Knispel}
\affiliation{II. Physikalisches Institut,
	Universit\"at zu K\"oln,
	Z\"ulpicher Stra\ss e 77,
	D-50937 K\"oln,
	Germany}

\author{Vasily~Tseplyaev}
\affiliation{Peter Gr\"{u}nberg Institut and Institute for Advanced Simulation, Forschungszentrum J\"{u}lich und JARA, 52425 J\"{u}lich, Germany}
\affiliation{Physics Department, RWTH-Aachen University, 52062 Aachen, Germany}

\author{Gustav Bihlmayer}
\affiliation{Peter Gr\"{u}nberg Institut and Institute for Advanced Simulation, Forschungszentrum J\"{u}lich und JARA, 52425 J\"{u}lich, Germany}
   
\author{Stefan Bl\"ugel}
\affiliation{Peter Gr\"{u}nberg Institut and Institute for Advanced Simulation, Forschungszentrum J\"{u}lich und JARA, 52425 J\"{u}lich, Germany}

\author{Thomas~Michely}
\affiliation{II. Physikalisches Institut,
	Universit\"at zu K\"oln,
	Z\"ulpicher Stra\ss e 77,
	D-50937 K\"oln,
	Germany}

\author{Jeison~Fischer}
\email{jfischer@ph2.uni-koeln.de}
\affiliation{II. Physikalisches Institut,
	Universit\"at zu K\"oln,
	Z\"ulpicher Stra\ss e 77,
	D-50937 K\"oln,
	Germany}

\date{\today}

\DeclareGraphicsExtensions{.pdf}

%\begin{tocentry}
%\includegraphics{Fig.s/toc.pdf}
%\end{tocentry}

\begin{abstract}
Ultrathin magnetic films on heavy metal substrates with strong spin-orbit coupling provide versatile platforms for exploring novel spin textures. So far, structurally open fcc(110) substrates remain largely terra incognita. Here, we stabilize a metastable, unreconstructed Ir(110)-$(1 \times 1)$ surface supporting two layers of Fe. Combining spin-polarized scanning tunneling microscopy and \textit{ab initio} calculations, we reveal a right-handed Néel-type spin spiral along the [$\overline{1}$10] crystallographic direction with a period of 1.27~nm as the magnetic ground state. Our analysis reveals this spiral is of the Yoshimori type, i.e., driven by frustrated Heisenberg interactions, with the Dzyaloshinskii-Moriya interaction determining its cycloidal nature and handedness.

\end{abstract}

% insert suggested PACES numbers in braces on next line

%Keywords: exchange frustration, Dzyaloshinskii-Moriya interaction, elliptical skyrmions, spin-polarized STM

%\maketitle must follow title, authors, abstract, \pacs, and \keywords
\maketitle
\newpage

\noindent

%%%%%%%%%%%%%%%%%%   INTRODUCTION      %%%%%%%%%%%%%%%%%%%%%%%

\textit{Introduction}: Ultrathin magnetic films on heavy transition-metal substrates with strong spin-orbit coupling (SOC) have catalyzed breakthroughs in modern condensed matter physics. Landmark studies have revealed interface-induced Dzyaloshinskii-Moriya interactions (DMI), first demonstrated in a monolayer (ML) of Mn on W(110)~\cite{Bode2007}, have enabled the first observation of atomic scale skyrmion lattices found in a single ML of Fe/Ir(111)~\cite{Heinze2011}, and the first realization of isolated atomic-scale skyrmions in a PdFe bilayer on Ir(111)~\cite{Romming2013}. In general, one can say that these systems provide a rich landscape for exploring complex, exotic spin textures~\cite{Ferriani2008, Yoshida2012, Menzel2012, Herve2018, Romming2018, Kronlein2018, Spethmann2020} with nonzero vector or scalar spin-chiralities, topological phenomena, and their manipulation~\cite{Romming2013,Krause2016} driven by the interplay of reduced dimensionality, surface orientation, stacking sequence, magnetic Heisenberg and beyond-Heisenberg exchange interactions, and Dzyaloshinskii-Moriya interaction, due to substantial spin-orbit effects in structure inversion-asymmetric environments. Since these platforms are frequently replicated as multilayers~\cite{Moreau-Luchaire2016}, they have not only opened new avenues in the study of magnetism, but have also enabled potential applications in spintronics and information technologies~\cite{Sampaio2013,Song2020,Yang2021,Everschor-Sitte2024}.

Among the ultrathin magnetic film systems, Fe/Ir occupies a distinctive role. It has revealed the crucial role of interactions beyond the Heisenberg model, including four-spin~\cite{Heinze2011} and four-spin-three-site interactions, particularly in bilayer hcp-Rh/Fe/Ir(111)~\cite{Romming2018} and isoelectronic Rh(111) substrates~\cite{Kronlein2018}. These interactions give rise to exotic spin configurations, such as a ground state square skyrmion lattice or an antiferromagnetic (AFM) up-up-down-down state. Besides studies on (111)-oriented interfaces, (100)-oriented Ir interfaces have also been investigated. For instance, the occurrence of an atomic-scale chiral spin spiral in finite individual bi-atomic Fe chains on the $(5\times 1)$-Ir(001) surface~\cite{Menzel2012} has been demonstrated, showcasing the diversity of spin textures across different orientations of Ir substrates. It can be summarized that the magnetic structures of Fe on the different Ir substrates can be understood by a competition of ferromagnetic (FM) Heisenberg exhange, beyond-Heisenberg and Dzyaloshinskii-Moriya interaction. 

Despite the impact of the Ir substrate to exciting spin textures, studies focused on the specific Ir(110) substrate are currently lacking. In general, fcc(110) surfaces are of particular interest because their open structure facilitates functionalization and enables tuning of the electronic and magnetic properties in magnetic films. Owing to their C$_{2v}$ symmetry, fcc(110) surfaces can also develop anisotropic interfacial DMI~\cite{Hoffmann2017,Camosi2017,Liu2021}, which is essential for stabilizing elliptical skyrmions or antiskyrmions~\cite{Hoffmann2017}. Together with the inherent anisotropic Heisenberg interaction, they may allow to minimize the skyrmion Hall angle or support the coexistence of skyrmions and antiskyrmions with equal energy, making them promising candidates for racetrack memory devices~\cite{Huang2017,Xia2020,Cheng2021,Hoffmann2021}.  

%%%%%%%%%%%%%%%%%%   EXPERIMENTS      %%%%%%%%%%%%%%%%%%%%%%%

\textit{Experimental results}: Despite the potential of fcc(110) surfaces, their preparation is complicated by the complex surface reconstructions observed in spin-orbit materials such as Ir, Pt, and Au~\cite{Koch1991,Schulz2000,Niehus1984,Gritsch1991,Robinson1983,Hoefner1998}. To overcome this challenge we prepared a metastable unreconstructed surface of Ir. Cooling Ir(110) from 1200~K in oxygen prevents its nanofacet-reconstruction~\cite{Koch1991,Schulz2000} through O-adsorption. The adsorbed oxygen is subsequently titrated away with H at 470~K. The resulting Ir(110)-$(1\times1)$ remains thermally stable up to 520~K. The measurements were performed in an ultrahigh vacuum scanning tunneling microscope (STM) operating at $4.2$\,K and equipped with a vector magnetic field up to 9\,T (out of plane) and 2\,T in any direction. Spin-polarized STM tips were prepared by coating a W tip with Fe. For more details on experimental methods and sample preparation, see Note 1 in the Supplemental Material (SM)~\cite{sm}.

%Fig. 1
\begin{figure}[h!]
\centering
\includegraphics[width=0.4\textwidth]{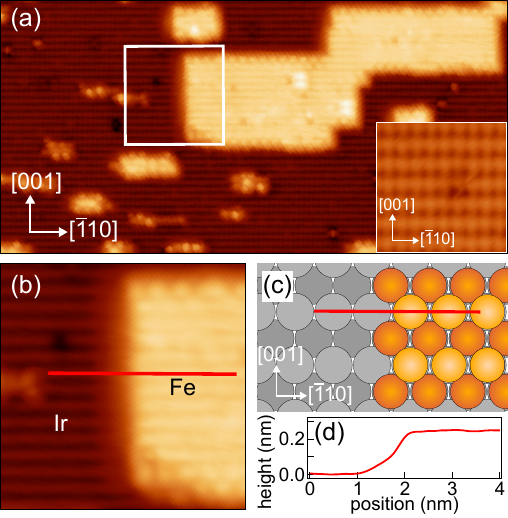}
\caption{2\,ML thick Fe islands on unreconstructed Ir(110)-$(1\times1)$.
(a) Constant-current STM image taken with a non-magnetic tip after deposition of 0.4\,ML Fe on Ir(110)-$(1\times1)$. $V_\mathrm{b} = 100$\,mV, $I_\mathrm{set} = 1$\,nA, and image size 28\,nm\,$\times$\,28\,nm. Inset: atomically resolved Ir(110)-$(1\times1)$. $V_\mathrm{b} = 50$\,mV, $I_\mathrm{set} = 30$\,nA, and image size 3\,nm\,$\times$\,3\,nm.
(b) Zoomed area indicated in (a) by white square.
(c) Top view ball model of Ir(110)-$(1\times1)$ with pseudomorphic Fe island on the right (Ir: grey, Fe: orange). 
(d) STM height profile along red line in (b).
}
\label{system}
\end{figure}

An STM topograph of Ir(110)-$(1 \times 1)$ after deposition of 0.4~ML Fe is shown in Fig.~\ref{system}(a). Fe deposition results in the formation of Fe islands with a height of 2~ML [see Fig.~\ref{system}(b)-(d)] featuring lateral sizes on the order of 10\,nm along with smaller Fe islands.
The densely packed atomic rows of bare Ir(110)-$(1 \times 1)$ along [$\overline{1}$10] align with those of the 2~ML Fe island [compare Fig.~\ref{system}(b) and (c)], consistent with pseudomorphic growth of the first two Fe layers. The apparent height of the Fe islands derived from profiles such as in Fig.~\ref{system}(d), is 0.26\,nm, closely matching expectations for pseudomorphic growth and slightly below the 0.27~nm height observed for two Ir layers on Ir(110).

%Fig. 2
\begin{figure*}[t!]
\centering
\includegraphics[width=\textwidth]{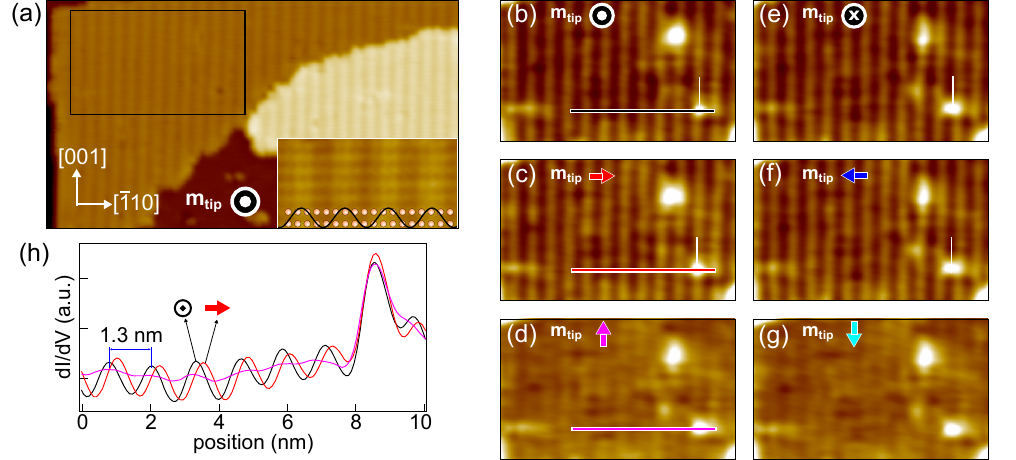}
\caption{Spin-polarized STM data obtained using a magnetically soft Fe-coated W tip with magnetization \textbf{m}$_\mathrm{tip}$ whose indicated direction is the one of the external field $B = 1$\,T. (a) Spin-polarized STM topograph of 2\,ML Fe on Ir(110) with a substrate step crossing from lower left to upper right. Only the area in the lower center is not covered by Fe. A magnetic wave pattern of the out-of-plane magnetization direction with wave vector in [$\overline{1}$10]-direction is visible in 2\,ML Fe areas. $V_\mathrm{b} = 100$\,mV, $I_\mathrm{set} = 5$\,nA, and image size is $38$\,nm\,$\times$\,$20$\,nm. Inset: spin- and atomically-resolved STM topograph. Circles are superimposed at locations of atomic intensity maxima in the lower part of the image. Magnetic wave pattern schematically represented by a sinusoidal curve is incommensurate with atomic positions. $V_\mathrm{b} = 50$\,mV, $I_\mathrm{set} = 1$\,nA, and image size is 5.0\,nm\,$\times$\,2.5\,nm. (b)-(g) Spin-polarized $\mathrm{d}I/\mathrm{d}V$ maps taken at indicated \textbf{m}$_{\mathrm{tip}}$ defined by external field. Magnetic contrast is observed for \textbf{m}$_\mathrm{tip}$ normal to the surface in (b) and (e) as well as along or opposite to [$\overline{1}10$] in (c) and (f). The magnetic contrast vanishes in (d) and (g) for \textbf{m}$_\mathrm{tip}$ along or opposite to [001], i.e., for \textbf{m}$_\mathrm{tip}$ normal to the pattern wave vector. $V_\mathrm{b} = 100$ mV, $I_\mathrm{set} = 5$\,nA, and image size is 16\,nm\,$\times$\,10\,nm. (h) Averaged $\mathrm{d}I/\mathrm{d}V$ profiles for colored line-like rectangles in (b)-(d) (colors matched). The profiles are taken with respect to a defect marked by vertical white lines in (b)-(d) as reference.
}
\label{sp_pristine_Fe}
\end{figure*}

Spin-polarized STM measurements were conducted with a magnetically soft Fe-coated W tip (for details see~\cite{sm}). The tip magnetization, \textbf{m}$_\mathrm{tip}$, aligns to the direction of an external magnetic field $\textbf{B}$, enables the decoding of the spin texture within the Fe film, as shown in Fig.~\ref{sp_pristine_Fe}. The topograph of Fig.~\ref{sp_pristine_Fe}(a) with $\textbf{B}$ and thus \textbf{m}$_\mathrm{tip}$ normal to the surface reveals a magnetic wave pattern in the Fe film with wave vector along [$\overline{1}$10]. The differential conductance ($\mathrm{d}I/\mathrm{d}V$) map in Fig.~\ref{sp_pristine_Fe}(b), taken within the black rectangle depicted in Fig.~\ref{sp_pristine_Fe}(a), displays the same spin-dependent modulation pattern. The brighter wave crests and darker valleys can be straightforwardly interpreted to result from the Fe spin component parallel or antiparallel to the tip magnetization, according to $\mathrm{d}I/\mathrm{d}V \propto \textbf{m}_\mathrm{s} \cdot \textbf{m}_\mathrm{tip}$ with $\textbf{m}_\mathrm{s}$ being the local sample magnetization~\cite{Wortmann2001}. Additional information on bias-dependent magnetic contrast is provided in Note 2 in the SM~\cite{sm}. The average wavelength $\lambda$ of the magnetic pattern is $\lambda = 1.27 \pm 0.02$\,nm, fairly independent of island size or shape. With $a_{\mathrm{nn}}= 0.2715$\,nm being the atomic spacing along the wave vector direction of the magnetic pattern, the wavelength $\lambda = 4.69\,a_{\mathrm{nn}}$ is incommensurate with respect to the crystal lattice periodicity. Support for the incommensurability of the magnetic pattern with the atomic lattice is given by the spin- and atomically-resolved inset of Fig.~\ref{sp_pristine_Fe}(a), as the spin contrast maxima do not match to a specific position in the atomic lattice defined by the bright protrusions.

\begin{figure}
\includegraphics[width=.4\textwidth]{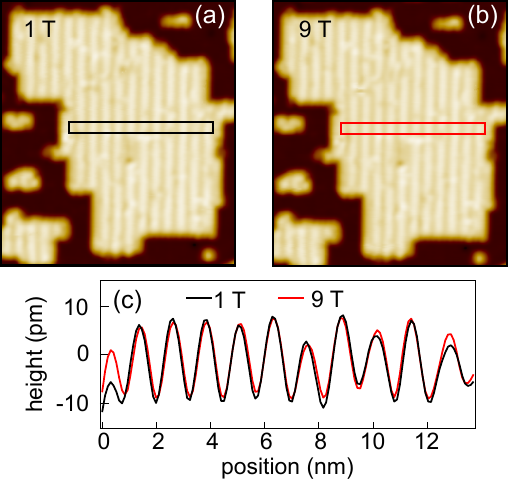}
\caption{Stability of frustated spin spiral against high magnetic field.
Spin-polarized STM images of 2\,ML Fe islands on Ir(110) at an external out-of-plane magnetic field of 1~T (a) and 9~T (b)  ($V_\mathrm{b} = 100$\,mV, $I_\mathrm{set} = 1$\,nA, 23\,nm\,$\times$\,26\,nm, and $T=4.2$\,K).
(c) Comparison of line profiles averaged over lines within the rectangle in (a) and (b). The modulation pattern is unchanged and sinusoidal behaviour is present in both cases.
}
\label{9T}
\end{figure}

To verify the stability of the magnetic pattern, we ramped the out-of-plane field from 1\,T to 9\,T, as shown in Fig.~\ref{9T}(a) and (b). The wave pattern period and intensity remains unchanged up to 9~T, which is even more evident when examining the line profiles in Fig.~\ref{9T}(c). This reveals that $\textbf{m}_\mathrm{s}$ is unaffected by the field, characterizing a hard magnetic sample texture. 

To fully characterize the magnetic texture, we probe other spin components by aligning the tip magnetization direction (${\textbf{m}}_\mathrm{tip}$) along the two orthogonal in-plane directions of the sample using a triple-axes vector magnet~\cite{Meckler2009}. Fig.~\ref{sp_pristine_Fe}(c)-(g) show $\mathrm{d}I/\mathrm{d}V$ maps obtained with $B = 1$\,T in the following directions: in-plane along the [$\overline{1}$10]-direction in (c), along the [001]-direction in (d), out-of-plane along the [$\overline{1}\overline{1}$0]-direction in (e), along the [1$\overline{1}$0]-direction in (f), and finally along the [00$\overline{1}$]-direction in (g). The contrast inversion of (e) with respect to (b) -- easy to spot with the help of the vertical white line marking a defect, (bright bump, marked by white vertical line) unaffected by the tip magnetization or external magnetic field -- confirms that \textbf{m}$_\mathrm{tip}$ follows the direction of the external $\textbf{B}$. It is obvious from the $\mathrm{d}I/\mathrm{d}V$ maps, that besides spin components normal to the surface [Fig.~\ref{sp_pristine_Fe}(b) and (e)], the texture displays also spin components parallel to the wave vector of the pattern [Fig.~\ref{sp_pristine_Fe}(c) and (f)], but no in-plane spin components normal to the wave vector [Fig.~\ref{sp_pristine_Fe}(d) and (g)]. These spin components are consistent with a flat Néel or cycloidal spiral, but inconsistent with a Bloch spiral.

The handedness of the Néel spiral is evaluated analyzing $\mathrm{d}I/\mathrm{d}V$ line profiles represented in Fig.~\ref{sp_pristine_Fe}(h). The line profiles are averages along the lines in Fig.~\ref{sp_pristine_Fe}(b)-(d). They are aligned with the help of the  defect. Turning the tip magnetization from normal [black curve, (b)] into the surface along the wave vector of the magnetic pattern [red curve (c)] leads to a shift of the wave pattern by about $\frac{1}{4} \lambda$ into [$\overline{1}$10] as visible in Fig.~\ref{sp_pristine_Fe}(h), defining a clockwise or right-handed rotation of the spin texture ($\color{gray}\downarrow$~$\color{blue}{\leftarrow}$~$\color{black}\uparrow$~$\color{red}\rightarrow$~$\color{gray}\downarrow$). Similar analyses for other islands unequivocally supports a unique rotational sense. To conclude, 2\,ML Fe on Ir(110) displays a clockwise rotating cycloidal or Néel-type spin spiral with a wavelength of $\lambda = 4.69\,a_{\mathrm{nn}}$. It is incommensurate to the underlying crystal lattice and magnetically hard with no changes in spin texture up to 9~T.

%%%%%%%%%%%%%%%%%%   THEORY   %%%%%%%%%%%%%%%%%%%%%%%

\textit{Theoretical results}: To investigate the physical origin of the robust spin spiral, we performed vector-spin density functional theory (DFT) calculations using the film version of the full-potential linearized augmented plane wave method~\cite{wimmer1981flapw} (FLAPW) as implemented in the FLEUR code~\cite{fleur2023} to analyze their energetics in a structurally optimized pseudomorphic, two-layer Fe film on an 11-layer Ir substrate modeling the experimental sample. The local density approximation (LDA)~\cite{Vosko:80} was applied, resulting in an optimized lattice constant of $a_0=0.382$~nm in good agreement to experiment (0.3839~nm). The structural optimization of the film includes the first four layers of the surface with relaxations of $-$19\%, $-$7\%, $-$1\% of the first, second and third layer in units of the interlayer distance in [110]-direction, $a_0/\sqrt{8}$. The calculations of  homogeneous spin spirals have been performed in the $p(1\times1)$ unit cell, exploiting the generalized Bloch theorem~\cite{Kurz:04} for calculations without SOC. The contribution of SOC to the spirals has been treated in first order perturbation theory~\cite{Heide:09}. 3840 $\bf k_{\|}$ points in the two-dimensional Brillouin zone have been used.

\begin{figure}[t!]
\includegraphics[width=0.4\textwidth]{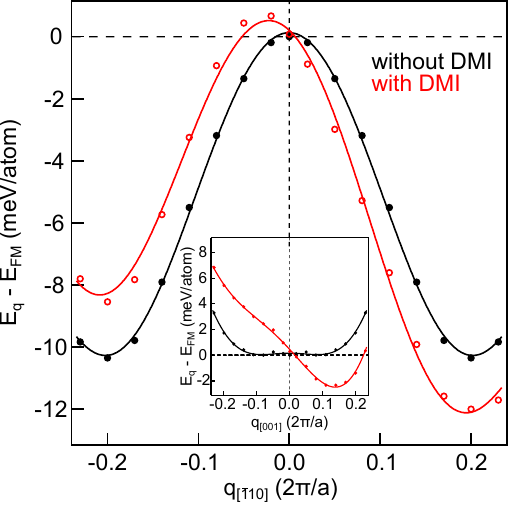}
\caption{DFT calculated energy dispersions $E(q)$ relative to the FM state of homogeneous, flat cycloidal spin spirals as a function of the wave vector $q$ for 2~ML Fe/Ir(110). 
Films with (red) and without (black) DMI along two orthogonal propagation directions: $[\overline{1}10]$ (main plot) and $[001]$ (inset). Positive (negative) $q$ indicates a right- (left-) handed rotational sense. Solid lines are polynomial fits to data point closely matching the points up to the Brillouin zone boundary (not shown).
}
\label{dft}
\end{figure}

The DFT results are summarized in Fig.~\ref{dft}. Calculations without SOC reveal a spin-spiral with a symmetric ($\mathbf{q} \to - \mathbf{q}$) dispersion featuring two deep energy minima ($<\!\!-10$~meV below the FM state) at finite, accidental, i.e., not symmetry determined values of $\pm q$, with  $q_{[\overline{1}10]} = 0.194 \left(\frac{2\pi}{a}\right)$, along the high-symmetry line $\left[\overline{1}10\right]$. This corresponds to an atomic-scale  spin-spiral period of $\lambda=1.39$~nm. The solution is an example of the well-known Yoshimori~spiral~\cite{Yoshimori1959} resulting from frustrated exchange, arising from competing FM and AFM Heisenberg interactions between different pairs of atoms~\cite{Phark2014}. When SOC is included, for a cycloidal N\'eel-type spiral, the associated DMI lifts the  energetic degeneracy by $\pm 2$~meV, the minima become asymmetric, favoring energetically a chiral ground-state N\'eel spiral with the positive $q$-value, i.e., a clockwise rotating spiral, though the DMI does not significantly alter the position of the energy minimum.  Along the $\left[001\right]$ direction, a different scenario is observed: Without SOC, no Yoshimori-spiral is observed, the FM state has the lowest energy, but the  DMI selects a positive $q_{[001]} = 0.14 \left(\frac{2\pi}{a}\right)$ corresponding to $\lambda_{[001]}=2.08$~nm, though at higher energy than for the $\left[\overline{1}10\right]$ direction. Consequently, a ground-state spin spiral stabilizing along $\left[001\right]$ is less favorable.

%%%%%%%%%%%%%%%%%%   SUMMARY / CONCLUSION     %%%%%%%%%%%%%%%%%%%%%%%

\textit{Discussion}: Comparing theory with experiment, we conclude: (i) The direction of the spiral agrees, (ii) the sign of the q-vector and thus the handedness of the spiral agrees, and (iii) there is a good agreement between the periods determined by theory and experiment, $\lambda=1.39$~nm versus $\lambda_\mathrm{exp}=1.27\pm0.02$\,nm. Turning to the analysis of the energy, we map the dispersion without SOC to a classical Heisenberg model  $E=-\sum_i J_{0i} S^2 \hat{\mathbf{e}}_0\cdot\hat{\mathbf{e}}_i$, with $\hat{\mathbf{e}}$ being the direction vectors of the local spin, we notice that the frustration of Heisenberg exchange between FM interaction of interplane nearest neighbor pairs and the AFM pair interaction of intraplane more-distant neighbors (for calculated exchange constants, see Note 3 in the SM~\cite{sm}) is five times stronger than the relativistic DMI. Thus, the frustrated exchange determines the energy and length scale of the spiral, but the DMI selects of the many possible spirals the clockwise rotating N\'eel-type one. Finally, the energy minimum is about 12~meV/Fe atom below the FM state, suggesting that  an external magnetic field of around $80$~T would be required to  unwind it into a skyrmion texture. This is consistent with the observed stability up to the highest applied field of $9$~T.

Mapping the energy dispersions around $q=0$ to a micromagnetic model where the strength of the Heisenberg exchange is expressed by the spin stiffness $A$ in $E(q) = A q^2$ and of the DMI by the spiralization $D$ in $E(q) = D q$, we find a strong asymmetry of the exchange interaction in different directions, $A_{[\overline{1}10]}=-19.8$~meV (minus sign implies, FM state is not stable), and  $A_{[001]} = -0.1$~meV, while the spiralization is fairly isotropic $D_{[\overline{1}10]}=-5.2$~meV/nm, and  $D_{[001]}=-5.0$~meV/nm, with the same handedness in both directions.

%Fig. 4

\textit{Summary}: In summary, we find that the magnetic ground state of pseudomorphic double layer Fe islands on unreconstructed Ir(110) is a frozen clockwise Néel-type atomic-scale cycloidal spin spiral along the [$\overline{1}$10]-direction with a wavelength of 1.27~nm. This spin spiral exhibits magnetic stiffness, remaining stable under external fields up to 9~T. DFT calculations show that, unlike (111) and (100) Ir substrates, the spin spiral here is of Yoshimori type, caused by frustrated Heisenberg exchange interactions, with the Dzyaloshinskii-Moriya interaction favoring a Néel over a Bloch spiral and breaking the chiral symmetry that determines its unique handedness. The exchange interaction is highly anisotropic, while the DMI is relatively isotropic, and no significant beyond-Heisenberg interactions were observed. These findings provide new insights into stabilizing atomic-scale spin spirals on low-symmetry substrates, highlighting the role of anisotropic magnetic interactions in determining the ground state. Future work could explore tuning spin textures through surface modifications, leveraging the openness of fcc(110) substrates.

\section{Acknowledgement}
V.T.\ and S.B.\ thank Markus Hoffmann for his support in carrying out the DFT calculations.  
We acknowledge funding from Deutsche Forschungsgemeinschaft (DFG) through CRC 1238 Grant No.\  277146847, projects B06 and C01). S.B.\ and J.F.\ acknowledge financial support from  DFG through projects BL~444/16-2 and FI~2624/1-1  within the SPP 2137 (Grant No.\ 462692705). 
S.B.\ acknowledges financial support from
the European Research Council (ERC) under the European
Union's Horizon2020 research and innovation program (Grant No.\ 856538, project “3D MAGiC”).
G.B.\ gratefully acknowledges computing time granted through JARA-HPC on the supercomputer JURECA at Forschungszentrum Jülich.

\bibliography{bib_FeIr110}
\end{document}

\title{Supplemental Material:\\
Atomic-scale Dzyaloshinskii-Moriya-modified Yoshimori spirals in Fe double layer on Ir(110)}

\author{Timo~Knispel}
\affiliation\UK

\author{Vasily~Tseplyaev}
\affiliation{Peter Gr\"{u}nberg Institut and Institute for Advanced Simulation, 
			Forschungszentrum J\"{u}lich und JARA, 52425 J\"{u}lich, Germany}
\affiliation{Physics Department, RWTH-Aachen University, 52062 Aachen, Germany}

\author{Gustav Bihlmayer}
\affiliation{Peter Gr\"{u}nberg Institut and Institute for Advanced Simulation, 
			Forschungszentrum J\"{u}lich und JARA, 52425 J\"{u}lich, Germany}
   
\author{Stefan Bl\"ugel}
\affiliation{Peter Gr\"{u}nberg Institut and Institute for Advanced Simulation, 
			Forschungszentrum J\"{u}lich und JARA, 52425 J\"{u}lich, Germany}

\author{Thomas~Michely}
\affiliation\UK

\author{Jeison~Fischer}
\email{jfischer@ph2.uni-koeln.de}
\affiliation\UK

\date{\today}
\maketitle
\vspace{-1cm}
\tableofcontents
\newpage
\section*{Supplementary Note 1: Experimental methods}
The experiments were performed in an ultra-high vacuum (UHV) system (base pressure $<$ 3$\times$10$^{-10}$~mbar) equipped with a sample preparation chamber and a scanning tunneling microscope (STM) operating at 4.2~K with superconducting magnets producing in the STM sample position a magnetic field of up to 9~T normal to the sample surface and 2~T in any direction.

\begin{figure}[h]
\centering
\includegraphics[]{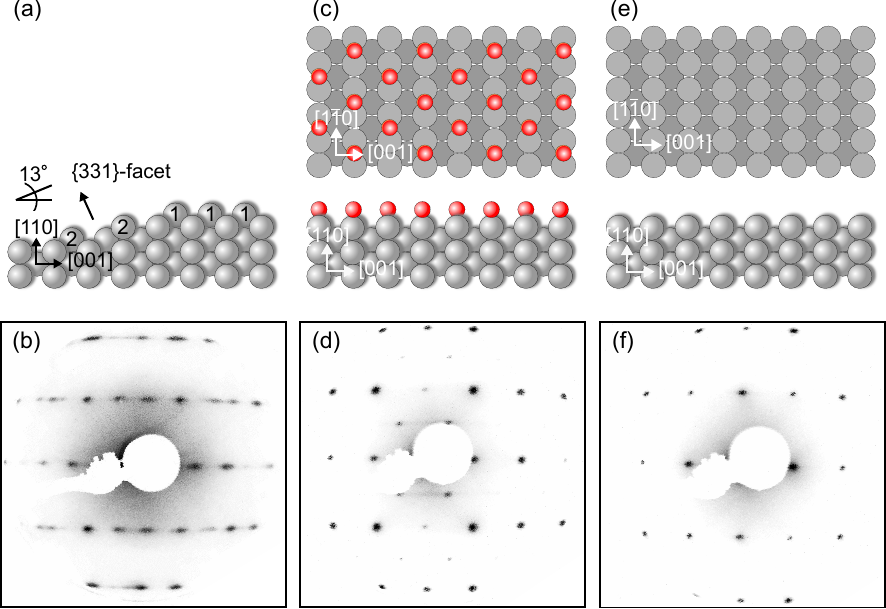}
\caption{Model and low energy electron diffraction (LEED) patterns of the preparation of unreconstructed Ir(110)-$(1 \times 1)$.
(a) Ball model (side view) of the (331)-facet of Ir(110) and (b) LEED pattern of the (331)-faceting of the Ir(110) surface. (c) ball model: top view (top) and side view (bottom) of the c$(2 \times 2)$-O superstructure and (d) LEED pattern of the Ir(110) with a c$(2 \times 2)$ pattern. (e) ball model: top view (top) and side view (bottom) of unreconstructed Ir(110) and (f) LEED image $(1 \times 1)$ corresponding LEED pattern. LEED energy $194$\,eV. Grey = Ir, red = O. 
}
\label{leed}
\end{figure}

Ir(110) was cleaned by cycles of 1~kV Ar-sputtering and annealing at 1500~K. After cool-down the sample displays the thermodynamically stable \{331\}-nanofacet reconstruction~\cite{Koch1991,Ney2002}. In order to establish clean unreconstructed Ir(110)-$(1 \times 1)$ we applied a two-step procedure. First, during cooling from 1200~K, the surface is exposed to an oxygen partial pressure of $1 \times 10^{-7}$\,mbar to prevent the formation of the nanofacet reconstruction. The result is oxygen covered Ir(110)-$(1 \times 1)$, as was first shown by Chan et al.~\cite{Chan1978}. Second, following the approach of Heinz et al.~\cite{Heinz1985} to prepare \textit{clean} unreconstructed Ir(001), chemisorbed oxygen was titrated through exposure to $1 \times 10^{-7}$\,mbar hydrogen at 470\,K until clean Ir(110)-$(1 \times 1)$ resulted. The corresponding low energy electron diffraction patterns and a model showing the different steps of preparation are given in the Figure~\ref{leed}.
Fe amounts ranging from 0.4\,ML to 1.4\,ML Fe were deposited by e-beam evaporation onto clean Ir(110)-$(1 \times 1)$ at 470~K. Here, ML denotes monolayer and 1\,ML corresponds to the surface atomic density of Ir(110)-$(1 \times 1)$, which is $9.6 \times 10^{18}$\,atoms/m$^2$.

For spin-polarized STM measurements, we used Fe-coated W. To obtain an Fe-coated W tip, an electrochemically etched W tungsten tip was heated in UHV up to 2400\,K for 2\,s and subsequently coated by 20 atomic layers of Fe. After deposition the Fe-coated W tip was annealed at 1000\,K for 5\,s~\cite{Phark2013}.

We detect the tunnel current $I$($V$) and the differential conductance ($\mathrm{d}I/\mathrm{d}V$) simultaneously using a lock-in technique with a modulation bias voltage of $20$~mV at a frequency $6.662$~kHz.

\clearpage
\section*{Supplementary Note 2: Energy-dependence of spin-polarized \MakeLowercase{d}\textit{I}/\MakeLowercase{d}\textit{V} of spin spiral in F\MakeLowercase{e} on I\MakeLowercase{r}(110)}

\begin{figure}[h]
%\centering
\includegraphics[width=.7\textwidth]{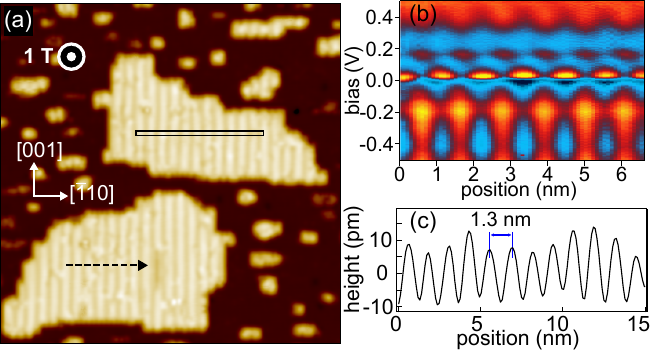}
\caption{(a) Constant-current STM image of 2\,ML Fe islands on Ir(110) at an external out-of-plane magnetic field of 1 T ($V_\mathrm{b} = 100$\,mV, $I_\mathrm{set} = 1$\,nA, and 30\,nm\,$\times$\,30\,nm). Magnetic modulation pattern parallel to the [001]-direction and with wave-vector aligned to the [$\overline{1}10$]-direction is visible. Bright and dark stripes represent local magnetization parallel and antiparallel to the out-of-plane tip magnetization.
(b) $\mathrm{d}I/\mathrm{d}V$ linescan along the dashed arrow shown in (a). The recorded $\mathrm{d}I/\mathrm{d}V$ signal is plotted as a function of bias voltage $V_\mathrm{b}$ and position $x$.
(c) Line profile averaged over lines within the rectangle in (a). The modulation pattern has a periodicity of 1.3~nm.
}
\label{linescan}
\end{figure}

\clearpage

\section*{Supplementary Note 3: Distant-dependent exchange parameters}

\begin{figure}[h]
\centering
\includegraphics[width=0.6\textwidth]{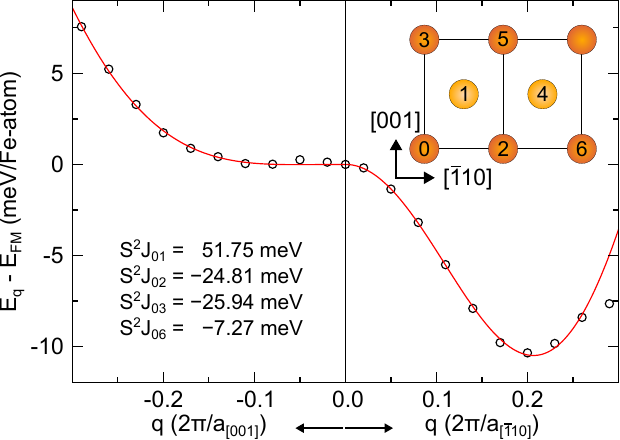}
\caption{Fitting (red line) of the energy dispersion of homogeneous spin spirals calculated by means of scalar-relativistic DFT for Fe/Ir(110) (black circles) as shown in Fig.~4 (main text). Negative (positive) sign of $q$ corresponds to $[001]$ and ($[\overline{1}10]$) directions. Ball model of the double layer Fe. Bright orange: top layer, dark orange: bottom layer. 
Nearest neighbor exchange constant is interplane and contributes equally, $J>0$, FM, in both directions. Next nearest neighbor is intraplanar and contribution is mostly AFM ($J<0$), but contribution differ in each direction. The resulting negative spin stiffness $A_{[\overline{1}10]}=-19.8$~meV is evidence of frustration in the system~\cite{Fischer2017}. Along the $[001]$ spin stiffness is also (slightly) negative, $A_{[001]}=-0.1$~meV, but the shallow energy minimum indicates the spiral may not be realized as the spins have to rotate in a plane containing the magnetically hard [001] direction ($E_{[001]}- E_{[110]} = 1.18$~meV/Fe and $E_{[\overline{1}10]}- E_{[110]} = 0.32$~meV/Fe).
}
\label{fig_Jij_parameters}
\end{figure}

\newpage

\bibliography{bib_FeIr110}